# High Frequency Acousto-electric Single Photon Source

C L Foden[1], V I Talyanskii[2], G J Milburn[3], M L Leadbeater[1] and M Pepper[1,2]

[1]Cambridge Research Laboratory, Toshiba Research Europe Ltd, 260 Cambridge Science Park, Milton Road, Cambridge CB4 0WE UK

[2]Cavendish Laboratory, University of Cambridge, Madingley Road, Cambridge CB3 0HE UK

[3]Department of Physics, University of Queensland, St. Lucia 4072, Australia

## Abstract

We propose a single optical photon source for quantum cryptography based on the acousto-electric effect. Surface acoustic waves (SAWs) propagating through a quasi-one-dimensional channel have been shown to produce packets of electrons which reside in the SAW minima and travel at the velocity of sound. In our scheme these electron packets are injected into a p-type region, resulting in photon emission. Since the number of electrons in each packet can be controlled down to a single electron, a stream of single (or N) photon states, with a creation time strongly correlated with the driving acoustic field, should be generated.

Single photon states at optical frequencies provide an ideal resource for quantum communication and information processing[1]. The recent demonstration of teleportation[2] using entangled photon pairs generated by parametric down-conversion is an example. In addition, there are proposals and experimental schemes to use single photons for quantum computation[3,4], but the most important and experimentally relevant application of single photon sources is in secure optical communications using quantum cryptographic protocols[5]. No true, unconditional, single photon sources exist at optical frequencies so other sources are used. Parametric down-conversion is used in conditional schemes where detection of one member of the pair determines the space-time location of the other. Highly attenuated coherent light approximates an unconditional source and is used for current optical implementations of quantum cryptography. However, weak coherent states may contain any number of photons, with an average of less than one, rendering present quantum cryptographic implementations insecure[6].

At the single photon level, the statistical fluctuations of an electromagnetic (EM) field can be described in terms of the variation and correlation in the detection times of individual photons[7]. The distribution of detection events in a counting time for a coherent field is described by a Poissonian distribution about the mean. A thermal field has super-Poissonian fluctuations that correspond to the detection events tending to occur together or 'bunched'. Light with sub-Poissonian noise is non-classical, and has detection events with a more regular spacing, the photons are 'anti-bunched'. Fluctuations cannot be reduced below the classical, or shot noise, level using any conventional optical components, but can be reduced by methods known as squeezing[8]. However, this has so far produced only small effects and seems to have limited application. The absolute limit for sub-Poissonian light is equally spaced detection events, corresponding to highly correlated emission events in the source. For such light, the intensity fluctuations are zero, although the phase fluctuations are maximal.

Suppression of noise on a current of electrons is relatively easy to achieve since the strong Coulomb interaction between electrons assists in regulating electron flow and controlling current fluctuations. The noise on a macroscopic current can be reduced below the shot noise level by simply passing the current through a resistor or current junction[9]. In mesoscopic devices, where the electron wavelength is comparable to the dimensions of the device, the electronic properties are determined by quantum mechanics. Recent advances in this field have meant that Coulomb blockade effect devices, zero-dimensional systems through which the flow of electrons is regulated at the single particle level, can be reliably and routinely made[10]. Complete control over the flow of the electrons is sought by metrologists for current and capacitance standards[11].

Control of electron flow is the key to creating both a current standard and a single photon source in the solid state. Indeed the first proposal for a source of heralded photon states[12] employed the single electron-hole turnstile, based on a current standard system[13,14]. This proposal was realised recently[15]. We propose a method of producing single photon states utilizing the acousto-electric effect. Surface acoustic waves (SAWs) propagating on a piezoelectric substrate are accompanied by a travelling wave of electrostatic potential within the crystal. Free charge, for example a two-dimensional electron gas (2DEG), interacts with the potential[16], and an acousto-electric drag current is created. Previous work has shown that a current quantized to a high degree of accuracy is obtained by further confinement of the 2D acousto-electric current with a split gate or quantum point contact (QPC)[17].

In the single photon source we propose, a SAW is created by application of a microwave signal to a series of interdigitated transducers[17], and propagates towards a lateral n-i-p junction (Fig. 1). A lateral junction could be fabricated through the use of back gates to induce the electron and hole gases, by ion implantation or by amphoteric doping of a non-planar substrate[18]. Within the 2DEG, the SAW electrostatic potential can be screened by the mobile charge but, in the vicinity of the QPC, where the charge density is lower, the SAW amplitude increases and can be sufficient to trap electrons in the SAW potential minima and carry them through the QPC. Propagation of a sufficiently powerful SAW through a QPC, biased beyond the conductance pinch-off, confines the electrons to a series of quantum dots, of a size determined by the SAW wavelength (of order one micron) and the channel geometry, moving at the velocity of sound (of order $10^3$ ms$^{-1}$ in III-V semiconductor materials). The small wavelength of the SAW means that the Coulomb energy of the dot formed in each SAW minimum is large enough to ensure that the number of electrons per dot is constant. Although the dot size is a continuous function of the gate voltage which defines the QPC, the number of electrons per dot has a step-like dependence on the gate voltage due to the gaps in the energy spectrum induced by the Coulomb interaction. Therefore the current induced by the SAW through the one-dimensional (1D) channel is also a step-like function of gate voltage with the value of the current on the plateaux quantized as *I=Nef*, where N is the number of electrons per quantum dot and the SAW frequency, *f*, is of order GHz. Quantization of the acousto-electric current induced by the SAW in a 1D channel has been observed[17]. The SAW pump used in that work is shown schematically in Fig. 2 along with a typical experimental dependence of the induced current on the gate voltage. It differs from the pump shown in Fig. 1 in that here the electrons are transferred between two n-type regions formed in a GaAs/AlGaAs heterostructure. Quantized plateaux in the current are seen, corresponding to the transfer of 1,2,3… electrons per SAW cycle. The first plateau is the flattest, and allows determination of the quantized current value with a precision of 50ppm[17]. Recent measurements[19] of the absolute accuracy of the current on this plateau indicate that the deviation of the experimental current from the exact value of *I=ef* is related to the plateau flatness and is of the same order as the precision. Thus SAW pumps with an operational frequency around 3GHz and an absolute accuracy around $10^{-4}$ are now available. Note that, as far as the single electron transport in the 1D channel is concerned, the set-ups in Figs. 1 and 2 are identical, therefore the techniques developed earlier can be used, and the same performance of the SAW electron pump may be anticipated. The accuracy of the current quantization, or accuracy of the electron injection rate, will influence the rate at which both missing and additional photon emission events occur.

Beyond the split gate in the single photon source (Fig. 1), the insulating channel connecting the n- and p-regions widens and the dots become wires along the minima of the SAW potential. The transition between the n- and p-regions must be sufficiently slowly varying in energy to ensure that the electrons continue to be confined by the SAW potential. Previous experiments have shown that a SAW potential of 30mV gives a good quantized current[17]. Assuming this SAW potential and a linear potential drop between the n- and p-regions of 1.5eV, this requirement can be satisfied with an i-region of length several tens of microns.

The quantized current is then injected into the p-region where the screening of the SAW potential by the mobile holes ensures that each electron is free to recombine radiatively with a hole, producing a photon with a wavelength determined by the bandgap of the semiconductor and the Fermi energy of the hole gas. The degree of screening can be calculated in terms of the 2D conductivity[16]. For our purposes it suffices to estimate the maximum possible redistribution of the holes under the action of the SAW potential. A 2D hole distribution of the form: $\rho_h = \rho_0 e^{-ikx}$, where $k$ is the SAW wavevector and $x$ is the direction of propagation of the SAW, gives rise to an electrostatic potential within the plane of the 2D hole gas (2DHG) of $\psi_h = \frac{\rho_h}{\varepsilon_0 k(\varepsilon+1)}$ where $\varepsilon_o$ is the vacuum permittivity and $\varepsilon$ is the dielectric constant of the material (12 in GaAs). By equating this potential to the SAW potential of 30mV, the charge density required to completely screen the SAW within the 2DHG is calculated to be $10^{10}$ cm$^{-2}$. Since a typical 2D carrier density is generally two orders of magnitude higher than this, it is a good approximation to neglect the SAW potential within both the n- and p-regions of the device.

The recombination process, to a first approximation, is a Poisson point process with a single rate $\gamma$. If N electrons are injected into the p-region, the state of the optical field in the output mode at time t after injection is

$$\rho(t) = \sum_{n=0}^{N} \binom{N}{n} e^{-n\gamma t}(1-e^{-\gamma t})^{(N-n)}|N-n\rangle\langle N-n| \qquad (1)$$

This will only produce a perfect number state if $\gamma t \gg 1$. The time-dependent probability for state $|m\rangle$ is then given, for $m \leq N$, by

$$p(m,t) = \binom{N}{m} e^{-(N-m)\gamma t}(1-e^{-\gamma t})^m \qquad (2)$$

It is therefore necessary to work in a regime such that, if T is the period between successive injection of electrons, $\gamma T \gg 1$. Note that operation of all suggested single photon sources, including our own, is limited by the uncertainty in the photon emission time, determined by the total recombination rate, and by missing photon emission events which may arise from non-radiative recombination processes. However, photoluminescence measurements of minority carrier recombination rates in high quality remotely doped GaAs quantum well samples have shown that radiative recombination dominates, and that the radiative recombination time, $\gamma^{-1}$, is of order 100ps[20]. Depending upon the level of accuracy required by the application[5], it may be necessary to increase either $\gamma$ or T (or both) in order to satisfy the above inequality. T can be made greater than one SAW period either by electrically gating the quasi-one-dimensional channel in synchronism with the SAW field, allowing a single electron to pass only every $M$ cycles of the SAW field or, similarly, by using a gated current Y-splitter[21] which allows an electron from the quantized current to enter the p-region only every $M$ cycles of the field (Fig. 3). (Note that reducing the SAW frequency is not a possibility as a longer wavelength SAW would not produce a well quantized current). The accuracy or acceptable rate of events when more than one photon is emitted per electron injection cycle limits, through Eq. 2, the maximum electron injection rate. For example, assuming a recombination time of 100ps and an accuracy requirement of less than one multi-photon emission event in $10^6$ (i.e. $(1-p(1,t)) \leq 10^{-6}$), Eq. 2 gives the maximum injection frequency to be 0.72GHz (or maximum injection current of 115pA). For a typical SAW frequency this would require the electron injection rate $T^{-1}$ to be reduced by a factor $M<10$. This accuracy condition also requires that the probability of the two-electron injection events should not exceed the same level ($10^{-6}$ in the above example).

Another proposal for a single photon source employing the acousto-electric effect has been given[22]. In this device, electron-hole pairs are generated optically and confined in wires, along the minima and maxima respectively, of the travelling potential wave created by the SAW. The device also comprises a special kind of dot, the function of which is to sequentially capture from the SAW potential, as the charged wires pass the dot, one of each type of particle which can then recombine. The main drawbacks of this device were discussed by the authors in the paper: operation relies on the quantum dot having certain postulated properties which may not be possible in practice and the capture process is a probabilistic one.

Both the electron-hole turnstile device mentioned earlier[12,15] and the photon source we propose here are based on previously proven techniques and well established technology. However, the operating frequency of all demonstrated single electron current sources based on the Coulomb blockade (including the electron-hole turnstile[15]) has been limited, by the probabilistic and slow process of tunnelling, to a maximum value of around 10MHz[23]. This is more than two orders of magnitude lower than the frequency of demonstrated SAW single electron pumps[17]. Thus, on the basis of achieved repetition rates, it might be anticipated that the SAW source has the potential to produce a higher photon flux, although improvements in growth and fabrication techniques and device design could produce dots with comparable operating frequencies. Note that the projected performance for the electron-hole turnstile employed in Ref. 15 was estimated[12] to be 1% accuracy at an operating frequency of 1GHz.

It is also important to note that the current experimental operational temperature of the SAW electron pump is 5-7K, compared with 50mK for the electron turnstiles, such that closed cycle cryogenic systems can be used in practical applications of the source. However, since Coulomb blockade has already been demonstrated at higher temperatures, future turnstile devices may have similar operating temperatures to the SAW source. The SAW photon source can also be used for the generation of N-photon states, either by using several QPCs in parallel or simply by biasing the QPC such that the quantized current is *I=Nef*. Control over the generation time, repetition rate and mode structure of the N-photon pulses means that optical delay line techniques can be employed to create product states of many modes. Such states could then be entangled by a variety of linear and nonlinear optical processes for testing predictions of quantum information theory.

As a source for quantum cryptography it is not essential that the proposed device produce a single photon for every injected electron. The protocols currently used are not dependent on detecting every photon sent from source to receiver because only photons received are used to establish the shared key, although if too many photons are lost or fail to be sent, the bit rate for the channel may be unacceptably low. Furthermore, failure to generate a photon in every cycle, or failure to detect every photon that is generated, will not prevent the observation of a very strong correlation between the SAW frequency and the detected photons. Thus experiments which demonstrate a strong temporal correlation between detected photons and the SAW pump are still possible even for low recombination rates and low detector efficiencies.

In summary, we have proposed a new method to generate single photon states at optical frequencies. It is hoped that the scheme is flexible and fast enough (in terms of repetition rates) to meet the requirements of quantum optical cryptography[5]. This source could also be used for fundamental tests in quantum mechanics and could provide a brightness standard.

C L Foden would like to thank D M Whittaker, C E Norman and S Egusa for useful discussions.

**Figure Captions**

Figure 1 a) schematic of the single photon source b) the conduction (CB) and valence band (VB) edge profiles across the n-i-p junction and recombination.

Figure 2 a) pictorial representation of the mechanism responsible for the plateaux in acousto-electric current, electrons are captured from the 2DEG by the SAW potential minima and move in packets through the split gate b) the acousto-electric current as a function of split gate voltage showing the plateaux in current to occur at integer multiples of the SAW frequency.

Figure 3 the electron injection rate to the p-region may be reduced by using a current Y-splitter with, inset, the voltage applied to the gate determines which branch of the Y-splitter is taken.

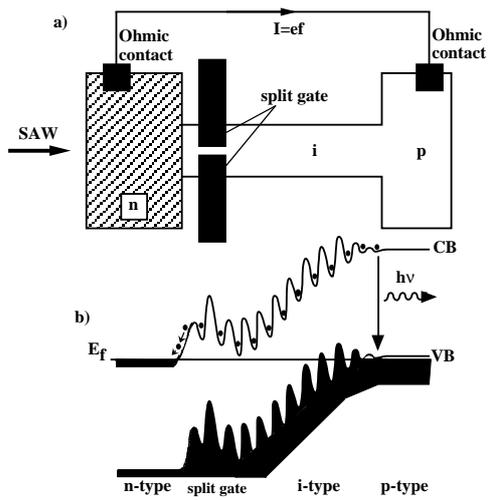

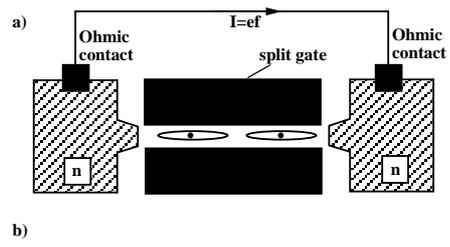

Figure 1

Figure 2

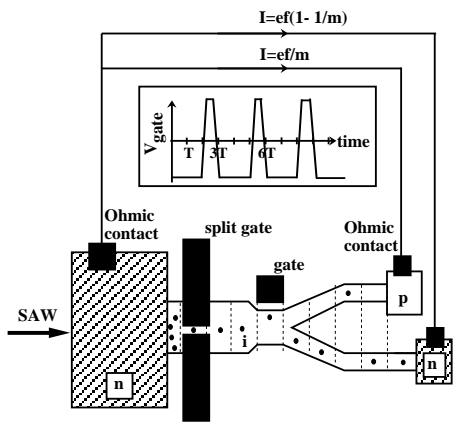

Figure 3